\documentclass[10pt,conference]{IEEEtran}

\usepackage[T1]{fontenc}
\usepackage[utf8]{inputenc}
\DeclareFontShape{T1}{ptm}{m}{scit}{<->ssub*ptm/m/sc}{}
\DeclareFontShape{T1}{ptm}{b}{scit}{<->ssub*ptm/b/sc}{}
\usepackage{graphicx}
\usepackage{booktabs}
\usepackage{amsmath}
\usepackage{xspace}
\usepackage{xcolor}
\usepackage[hidelinks]{hyperref}
\usepackage{orcidlink}
\usepackage[capitalise]{cleveref}
\definecolor{win}{RGB}{0,110,0}
\newcommand{\win}[1]{\textcolor{win}{\textbf{#1}}}


\newcommand{\sys}{\textsc{Inquest}\xspace}
\newcommand{\retr}{the retrieval layer\xspace}  
\newcommand{\nInstances}{38{,}001\xspace}   
\newcommand{\nScored}{37{,}900\xspace}      
\newcommand{\nProjects}{27\xspace}
\newcommand{\nHardPct}{49\%\xspace}         
\newcommand{\nMemoryPool}{172{,}331\xspace} 
\newcommand{\nEval}{26{,}544\xspace}
\newcommand{\nEvalHard}{12{,}799\xspace}
\newcommand{\accOne}{68.6\%\xspace}
\newcommand{\accFive}{85.9\%\xspace}
\newcommand{\mrr}{0.761\xspace}
\newcommand{\hardOne}{52.8\%\xspace}
\newcommand{\hardFive}{73.6\%\xspace}
\newcommand{\modOne}{90.2\%\xspace}         
\newcommand{\nRegion}{24{,}850\xspace}      
\newcommand{\regOne}{53.8\%\xspace}
\newcommand{\regFive}{69.9\%\xspace}
\newcommand{\retrOne}{41.2\%\xspace}
\newcommand{\retrFive}{70.3\%\xspace}
\newcommand{\retrHardFive}{48.7\%\xspace}

\newcommand{\loopRegOne}{39.8\%\xspace}
\newcommand{\loopRegFive}{50.7\%\xspace}
\newcommand{\calls}{15.7\xspace}
\newcommand{\tokens}{106k\xspace}
\newcommand{\nSample}{1{,}000\xspace}
\newcommand{\nSampleHard}{483\xspace}
\newcommand{\firewallPct}{63\%\xspace}
\newcommand{\moduleRetr}{76.5\%\xspace}    
\newcommand{\moduleFreq}{49.3\%\xspace}    

\begin{document}

\title{\textsc{Inquest}: Log-Guided Root-Cause Code Localization with a Staged Agent}

\author{
\IEEEauthorblockN{Yuvraj Sehgal\,\orcidlink{0009-0002-1980-604X}}
\IEEEauthorblockA{\textit{Department of Computer Science}\\
\textit{Brock University}\\
St.\ Catharines, Ontario, Canada\\
ys19rk@brocku.ca}
\and[\hspace{0.6in}]
\IEEEauthorblockN{Sneh Patel\,\orcidlink{0000-0002-1985-8715}}
\IEEEauthorblockA{\textit{Department of Computer Science}\\
\textit{Brock University}\\
St.\ Catharines, Ontario, Canada\\
sp18oo@brocku.ca}
\and[\hfill\vspace{1em}\break\hspace*{\fill}]
\IEEEauthorblockN{Mahsa Panahandeh\,\orcidlink{0000-0002-6369-8982}}
\IEEEauthorblockA{\textit{School of Electrical Engineering}\\
\textit{and Computer Science}\\
\textit{University of Ottawa}\\
Ottawa, Ontario, Canada\\
mpanahan@uottawa.ca}
\and[\hspace{0.6in}]
\IEEEauthorblockN{Naser Ezzati-Jivan\,\orcidlink{0000-0003-1435-6297}}
\IEEEauthorblockA{\textit{Department of Computer Science}\\
\textit{Brock University}\\
St.\ Catharines, Ontario, Canada\\
nezzatijivan@brocku.ca}
}

\maketitle

\begin{abstract}
When a production issue is reported, engineers get a description and some logs, but no failing test and no way to reproduce the fault. The logs show where the failure was noticed, not where it was caused: in 49\% of the real reports we study, the text names none of the files that the fix later changed. We built a benchmark for this setting from two public datasets and 27 open-source projects (Apache, Red Hat, MongoDB, MariaDB, Qt and Sakai). It holds 38,001 resolved issue reports with their pasted logs, the code as it was while the bug was still present, and the files the fix changed. We present \sys{}, an agent that finds the faulty code using a 27B open-weights model. Code, not the model, controls the procedure. It first gathers the evidence, including the files fixed in similar past issues that were resolved before the report was filed. The model then makes only three bounded decisions: which files to open, what role each file plays in the failure, and how to rank them. On 26,544 evaluation reports, \sys{} ranks a changed file first for 68.6\% and in the top five for 85.9\%, names the correct module for 90.2\%, and ranks a changed function first for 53.8\%. On the reports that name no faulty file, it reaches 52.8\% and 73.6\%. A free agent loop with the same model and tools finds files equally well, but names the faulty function 14 points less often. On a 1,000-report sample, \sys{} beats our reimplementations of two published agents by 7 and 20 points of top-1 accuracy (4 and 16 counting only the reports those agents answered), and the same model used as a re-ranker by 6. We report the cost of every system and the results that go against us, and we release the benchmark, both systems and every run transcript.
\end{abstract}


\section{Introduction}
\label{sec:intro}

When a production issue is reported, an engineer receives two things: a
description of what went wrong, and logs. There is no failing test, no
debugger session, and often no way to reproduce the problem. From this
evidence alone the engineer must find the code that caused it. This
paper is about doing that step automatically: given an issue report,
its logs and a source snapshot, produce the affected module, a ranked
list of suspicious code regions, and the evidence that connects them.

The core difficulty is that logs show where a failure was
\emph{detected}, not where it was \emph{caused}. Kafka issue KAFKA-6624
reports that a disk is wrongly marked offline, and its stack trace
points at \texttt{FileChannelImpl.force} and \texttt{FileRecords.flush},
the code that hit a closed file channel. The fix landed in
\texttt{LogManager.scala}, which allowed a log to be deleted while it
was still being flushed. Every file the trace names is innocent, and a
method that matches the report against the code it mentions, which is
what an engineer grepping the codebase does, starts from the wrong
files. In the benchmark we mined, \nHardPct{} of real reports name none
of the files the fix later touched.

Existing techniques approach this from two sides and stop short.
Classical fault localization ranks code using failing tests or
coverage, execution evidence that does not exist at report
time~\cite{just2014defects4j,rafi2024multiagent}. Log-based root cause
analysis in cloud operations identifies the faulty service or
component but stops above the level of code~\cite{wang2024rcagent}.
Agents that localize from an issue alone exist, but they are built and
evaluated on Python repositories~\cite{chen2025locagent,jiang2025cosil},
and they treat every issue as if it were the project's first. One source
of evidence is missing from all of them: the project's own resolved
history. Failures recur, and when an issue arrives, some past issue has
often failed in a similar way and its fix commit records exactly which
files were at fault. We call this \emph{incident memory}, and every use
of it here sits behind a temporal firewall that admits only incidents
resolved before the report was filed.

Studying this required a benchmark that did not exist. We mined one
from two public datasets and \nProjects{} git histories: \nInstances{}
real incidents from Apache and Red Hat projects, MongoDB, MariaDB, Qt
and Sakai, mainly in Java, Scala, C and C++, each pairing a report and
its logs with
the code at a revision where the defect is still present and the files
the fix later changed. The ground truth is the set of non-test files
the fix commit modified. That is a proxy for causation, not a
measurement of it: a fix also touches files hardened against
recurrence or refactored in passing, so where this paper says a system
localizes the root cause, the evidence is that it ranks the files the
fix changed.

We then built \sys{}, an agent that keeps the retrieval and the
incident memory as tools and reads code. It is not an agent of the
usual shape, one model choosing freely among tools until it decides to
stop. Its procedure is fixed in code: the evidence is gathered before
the model makes any decision, every candidate file is judged on its own with its
code on screen, code walks back from a file that only detects the
failure to the files that call it, and the model decides only three
things, which files to open, what role one file plays, and the order of
the files. On \nEval{} incidents \sys{} ranks a file the fix changed
first for \accOne{} and in the top five for \accFive{}, names the
correct module for \modOne{}, and names a function the fix changed
first for \regOne{}. An ablation that gives the same model the same
tools in an unconstrained loop finds files equally well and names the
faulty function fourteen points less often, which locates what the
structure is worth.

This paper makes three contributions.
\begin{itemize}
\item \textbf{A benchmark} of \nInstances{} issue--log--code--fix
  instances across \nProjects{} projects in four main languages, stratified
  by whether the report names any faulty file, with a per-project
  incident-memory pool of \nMemoryPool{} resolved issues and a datasheet
  for every filter (\cref{sec:benchmark}).
\item \textbf{\sys{}}, a staged localization agent in which code owns
  the procedure and the model decides inside bounded steps, with every
  decision recorded (\cref{sec:agent}). It returns the module, ranked
  file and function regions, and a causal chain whose steps cite the
  report's log events.
\item \textbf{An evaluation on the whole evaluation split}
  (\cref{sec:results}): \sys{} against the retrieval layer and against
  an unconstrained-loop ablation on \nEval{} instances, and against two
  reimplemented published agents and the same model used without tools
  on a \nSample-instance sample, with the cost of every system reported
  next to its accuracy.
\end{itemize}

The retrieval layer is deterministic, every agent tool is a pure
function of the snapshot, every run is recorded in full, and every
number in this paper traces to a result file through a released script.

\section{A Benchmark from Issues, Logs, and Fix Commits}
\label{sec:benchmark}

Evaluating this task needs issue descriptions, logs, source code and
ground-truth faulty code regions together. Existing resources supply
some of these but not the combination. Defects4J supplies failing
tests, which is exactly the execution evidence missing when an issue is
first reported~\cite{just2014defects4j}. Bench4BL pairs reports with
code and gold files at scale but does not select for reports that carry
logs~\cite{lee2018bench4bl}, and benchmarks built from pull requests
inherit solution text in the issue
thread~\cite{aleithan2024swebenchplus}. Pathidea assembled the closest
resource, a log-bearing set of Java issues~\cite{chen2022pathidea}.
What we add is a set selected for log-bearing reports across
\nProjects{} projects in Java, Scala, C and C++ (with some Python, Go
and Groovy), stratified by whether the
report names any faulty file, and shipped with a per-project pool of
past incidents so that history-based methods can be evaluated under a
stated temporal firewall. Throughout, an \emph{incident} is a resolved
issue report together with the log excerpts its reporter pasted into
it. These are the logs an engineer is handed at triage, not a raw
production log bundle; what that leaves out is discussed in
\cref{sec:discussion:difficulties}.

\subsection{Sources and construction}
\label{sec:benchmark:linkage}

We join two public datasets. World of Logs v2 records, for every public
Jira issue, whether its text contains log messages and an uncertainty
score for that extraction~\cite{wang2026wol}. The Public Jira Dataset v7
provides the metadata of 2.7 million issues from 16 trackers, with
creation and resolution times, resolution status and typed links
between issues~\cite{montgomery2022jira}. The two join exactly on the
issue key, and both are licensed CC-BY-4.0. From the trackers we took
every project whose fix commits can be linked to its issues through the
default branch of a public git repository: twenty Apache projects (Hive,
Hadoop, HBase, Ambari, Spark, Cassandra, Flink, Kafka and others), three
Red Hat projects (Infinispan, WildFly, Keycloak), MongoDB, MariaDB, Qt
(the \texttt{qtbase} module) and Sakai.

An issue enters the benchmark if its text contains logs according to
WoL with extraction uncertainty at most 0.05, if it is a bug resolved as
fixed, if it links to one to ten fix commits (reverts and merges
excluded), and if at least one of those commits modified a non-test
source file. The \emph{gold} of an instance is the set of non-test
source files the fix modified, capped at fifty files to exclude bulk
edits such as license-header changes, which Acc@K would otherwise
score as a free hit for every system. The \emph{snapshot} is the
repository at the parent of the fix commit: the code while the bug was
still present, with the fix never visible. The \emph{query} is the
issue summary and description in full. \Cref{tab:funnel} gives the
yield of every condition.

\begin{table}[t]
\caption{How the benchmark was mined, and what it holds. The scored set
drops the 101 instances whose gold is entirely test code, where a hit
would mean retrieving the test the fix edited rather than the defect;
they are kept in the release and excluded from every score.}
\label{tab:funnel}
\centering
\footnotesize
\begin{tabular}{@{}lr@{}}
\toprule
log-bearing fixed bugs, uncertainty $\le 0.05$ & 70{,}999 \\
\quad no fix commit found & $-$18{,}661 \\
\quad more than ten fix commits & $-$55 \\
\quad no non-test source file modified & $-$14{,}207 \\
\quad gold over fifty files (bulk edit) & $-$75 \\
\textbf{instances} & \textbf{\nInstances} \\
\quad gold entirely test code (scoring-time exclusion) & $-$101 \\
scored instances & \nScored \\
\midrule
projects / main languages & 27 / 4 \\
hard: report names no gold file & \nHardPct \\
report contains a stack trace & 30\% \\
single fix commit & 85\% \\
gold files per instance, median / p95 & 1 / 8 \\
fixes touching more than three files & 37\% \\
query length in characters, median / p95 & 902 / 6{,}577 \\
\bottomrule
\end{tabular}
\end{table}

The parent of the fix commit is a known-defective revision, not the
revision deployed when the report was filed; the two differ by whatever
landed in between. What we measure is therefore localization on a
revision that contains the defect, which is the setting this paper
studies, and not a replay of the exact tree an engineer faced at
triage. Report-time admissibility is enforced where we claim it: the
incident memory of \cref{sec:approach:memory} admits nothing resolved
after the report.

\subsection{Easy and hard instances}
\label{sec:benchmark:strata}

Logs often show where a failure was detected rather than where it
originated. We make this measurable. An instance
is \emph{easy} if the issue text names at least one gold file, for
example in a pasted stack trace, and \emph{hard} otherwise: the report
describes the failure, but no file the fix changed is named anywhere in
it. Across the benchmark \nHardPct{} of instances are hard; per project
the share runs from 22\% (PDFBox, whose reports usually carry a Java
trace) to 72\% (Groovy). The definition is file-based and therefore
favours Java, where a stack frame names the file; we also ship a second
label that counts a mention of a gold file's basename or of any function
it defines, under which the hard share falls, and we keep the file-based
one for stratification because a file name is what grep turns into a
candidate directly. Every result in this paper is reported on both
strata, and we also record whether the report contains a stack trace at
all.

\subsection{The incident memory pool}
\label{sec:benchmark:memory}

Our approach uses the project's resolved history as a signal
(\cref{sec:approach}). This history does not need logs: any fixed issue
whose fix commit can be identified tells us which files a past problem
ended in. We therefore build a separate \emph{memory pool} per project
from all of PJD using the same conditions without the log requirement,
\nMemoryPool{} issues in total. Each entry carries the issue's summary
and description, its creation and resolution times, its fix commits and
their gold files. PJD's typed links also give duplicate clusters,
groups of issues marked as duplicates of one another. Timestamps and
clusters are what keep the memory honest during evaluation
(\cref{sec:approach:firewall}).

\section{The Retrieval Layer: Log-to-Code Matching Plus Incident Memory}
\label{sec:approach}

The retrieval layer is a system in its own right in
\cref{sec:results:retrieval}, and it is the component \sys{} uses as a
tool, so its construction matters twice. It is lexical and
deterministic: no model, nothing trained, one CPU core, a median of under
0.2 seconds per issue on every project. Four signals each produce a ranked list over the
snapshot's non-test source files, and a weighted fusion merges them.

\textbf{Trace mentions (S1)} finds files named in the report itself:
file names with an extension (\texttt{KafkaChannel.java},
\texttt{row0mysql.cc:123}) and class names inside stack-trace frames,
ranked by where they first appear, since reporters paste the decisive
trace early. If no file is named, S1 falls back to bare file-name
stems. S1 is deliberately simple because it doubles as a baseline: it
is the automated form of an engineer grepping the report.
\textbf{Path retrieval (S2)} scores every file path against the report
with BM25 after splitting identifiers on case and underscores.
\textbf{Content retrieval (S3)} scores file contents, where the
identifiers, string literals and logging statements that produced the
reported lines live. To avoid a global index that leaks across time, S3
scores a per-instance pool, the files found by S1, the top 200 from S2
and the files proposed by memory, with term statistics computed inside
that pool only.

\label{sec:approach:memory}
\textbf{Incident memory (S4)} searches the project's pool
(\cref{sec:benchmark:memory}) for past incidents whose summary and
description read like the current report, with BM25 over an inverted
index. The twenty most similar admissible incidents vote: an incident
at similarity rank $r$ adds $1/r$ to every gold file of its own fix,
provided the file still exists in the current snapshot. Unlike S1 to
S3, S4 can point at a file that no log line and no trace ever names,
which is what the hard stratum requires.

\label{sec:approach:firewall}
\textbf{The memory firewall.} A memory signal evaluated on historical
data can cheat in ways that are invisible if they are not controlled.
A past incident is excluded from voting on a query when (1) it is the
query issue itself, (2) it belongs to the query's duplicate cluster,
(3) it shares a fix commit with the query, or (4) it was resolved after
the query was created. Rule 4 is strict and it costs real signal:
pooled over the benchmark it removes \firewallPct{} of the candidate
votes the index proposes. We accept the cost because the remaining
signal is one a deployed system would truly have had. One residual
channel remains and we state it: the index's corpus statistics are
computed once over the full pool, so the text of excluded incidents
still influences term weights; excluded incidents never vote.

\label{sec:approach:fusion}
\textbf{Fusion, conditioned on an observable regime.} The four lists
are merged by reciprocal rank fusion~\cite{cormack2009rrf}, file $f$ receiving
$\sum_i w_i / (60 + \mathrm{rank}_i(f))$. The weights $w_i$ are the only
tuned parameters. The right weights depend on the report: when the text
names files, S1 is nearly always right and should dominate; when it
names none, S1 is noise and memory matters more. The easy and hard
strata cannot drive this choice, because they are defined through the
gold answer, so we condition on a property visible at inference time,
whether S1 found an explicit file mention, and tune the weights per
regime by grid search over $\{0, 0.5, 1, 2\}$ on the development split
only (\cref{sec:setup:protocol}). The variant without memory is tuned
the same way, so the memory ablation compares two equally tuned
systems. The affected module is read off the file ranking at a fixed
per-project path depth, and every ranked file carries its provenance:
the signals that ranked it, the matching mentions or terms, and for S4
the keys of the past incidents that voted for it.

\section{\sys{}: a staged investigation}
\label{sec:agent}

\begin{figure*}[t]
\centering
\includegraphics[width=\textwidth]{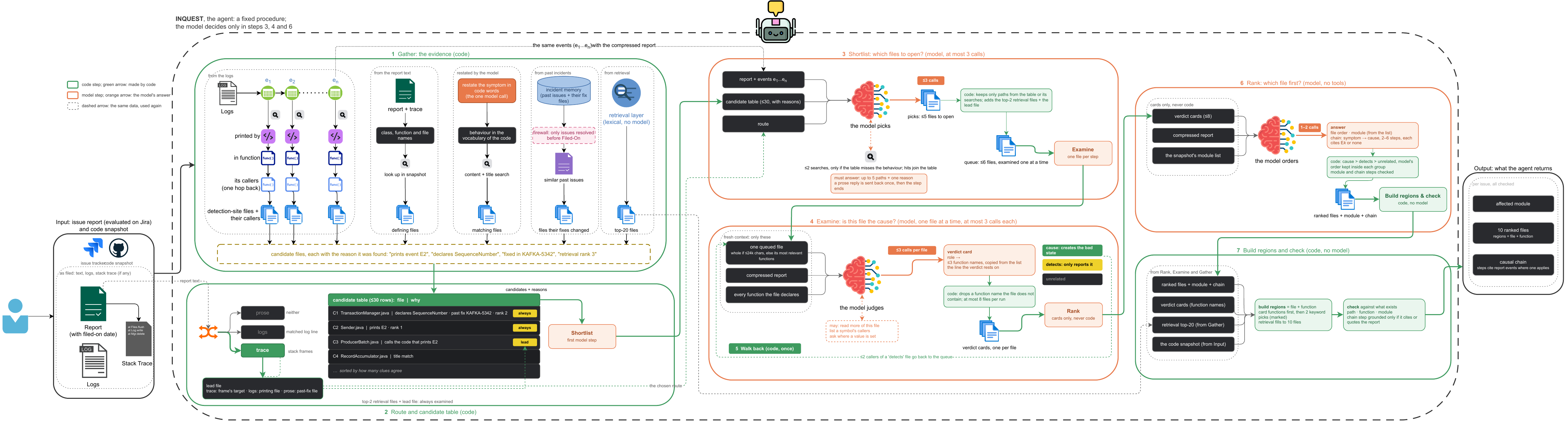}
\caption{\sys{}. The issue report and the code snapshot enter on the
left; the module, the ranked regions and the causal chain leave on the
right. Everything inside the dashed border is the agent: a fixed
procedure of seven steps. Code gathers the evidence and builds the
candidate table (1, 2). The model decides three times, each in a
bounded step with a fixed answer form: which files to open (3), what
role one file plays (4, one file per step, with that file on screen),
and the order of the files (6, from the verdict cards only). Code sends
the callers of a file that only detects the failure back to step 4,
once (5), and builds and checks the answer (7). Green boxes and arrows
are code; orange are the model; dashed arrows carry data that a later
step uses again.}
\label{fig:workflow}
\end{figure*}

\sys{} runs on one general-purpose 27B open-weights model, self-hosted,
at temperature zero with a fixed seed. Its inputs are the issue report
with its logs, the code snapshot, and the project's incident memory
behind the firewall of \cref{sec:approach:firewall}. Its outputs are
the affected module, a ranked list of code regions (file and
function), and a chain of steps from the reported
symptom to the named cause, each step citing a log event from the
report where one applies.

\subsection{Design principles}
\label{sec:agent:principles}

Localization agents are usually built as a loop: one model, a set of
tools, a turn budget, and the model deciding at every turn what to do
next and when to stop. \sys{} is built on four principles that take
those decisions away from the model. The evidence is gathered by code
before the model makes any decision, so no turn is spent searching for what the
report already names. Every candidate file is judged on its own, in a
fresh context with its code on screen, so a file's rank depends on what
its code does rather than on how often or how recently it was read.
The step from a file that only detects a failure to the files that
could have caused it is taken by code, not left to the model. And the
run ends when the procedure ends, with a submission the schema requires
to be complete. What this structure is worth is measured in
\cref{sec:results:structure} against an ablation that gives the same
model the same tools in an unconstrained loop.

\subsection{The staged workflow}
\label{sec:agent:stages}

\Cref{fig:workflow} shows the procedure. It is fixed; the model decides
only inside it, three times, and every decision is written to the
trace.

\textbf{1. Gather (code).} Code collects the evidence before the model
makes any decision. The logs in the report are grouped into events and labelled
\texttt{E1}, \texttt{E2}, \ldots; for each event the code finds the
statement that printed it and that statement's callers; every symbol in
the report is resolved to the file that defines it; the retrieval layer
of \cref{sec:approach} supplies its top twenty files; the incident
memory is queried, behind the firewall, for similar past incidents and
the files their fixes touched; a behaviour search and a title search run
over file content, the behaviour query being the one model call of this
step, a restatement of the symptom in code vocabulary.

\textbf{2. Route and table (code).} The report is routed by what it
contains: stack frames, else log lines that match a statement in the
code, else prose. Every file any channel named goes into one candidate
table of up to thirty rows, each with the reasons it is there
(``declares \texttt{SequenceNumber}'', ``fixed in past incident
KAFKA-5342'', ``prints log event E2''). The table is sorted by how many
kinds of evidence point at a file.

\textbf{3. Shortlist (model, at most three calls).} The model sees the
compressed report, the events and the table, may run at most two
searches if the table clearly misses the described behaviour, and must
answer with up to five paths to examine. Code keeps only paths that are
in the table or that the model's own searches returned, and always adds
the top two retrieval files and the route's lead file, whether or not
the model picked them; the queue is capped at six files, which leaves
two of the eight examination slots for walk-back.

\textbf{4. Examine (model, one file per step, at most three calls
each).} Each queued file is judged alone, in a fresh context that holds
only the report and that file: the whole file when it fits 24{,}000
characters, otherwise its most relevant functions whole, most relevant
first, until the budget is full, plus the list of every function the
file declares. The model may read another part of the same file, list a
symbol's callers, or ask where a value is set, and must answer with a
fixed-form verdict: the file's role (\emph{cause}: code here creates
the bad state; \emph{detects}: code here only reports or checks a bad
state made elsewhere; \emph{unrelated}), up to three function names
copied from the list, and the line the verdict rests on. A function name
the file does not contain is dropped, never repaired. At most eight
files are examined per run.

\textbf{5. Walk back (code, once).} For every file judged to
\emph{detect} the failure, code looks up the callers of the function the
verdict named and sends up to two caller files back to step 4.

\textbf{6. Rank (model, no tools).} The model sees the report and the
verdict cards, never raw code or tool output, and answers with the order
of the files, the affected module copied from the snapshot's module
list, and the causal chain, each step naming a log event id or none.
Code then applies one rule the prompt also states: files judged
\emph{cause} come before \emph{detects}, and those before
\emph{unrelated}, with the model's order kept inside each group.

\textbf{7. Build regions and check (code).} Regions are built per file
in rank order: the card's functions first (two for each of the first
three files, one after that), then two functions ranked by keyword
match against the report, marked as such. Every path must exist in the
snapshot, every function in its file, and the module in the module
list; every cited log event is checked against the report and marked as
grounded when it is one of the report's event ids or a line quoted from
the report word for word.
If fewer than ten files were examined, the retrieval ranking fills the
remaining file slots without reordering what the agent produced.

The whole run spends \calls{} model calls on average: one to restate
the symptom in code vocabulary during gathering, up to three to
shortlist, up to three per examined file, and one or two to rank. The
model is never asked to remember what it read in an earlier step,
because every step receives what it needs in full, and it is never
asked to decide when to stop, because the procedure stops.

\subsection{Reproducing an agent}
\label{sec:agent:repro}

Every tool is a pure function of the snapshot, the memory pool and the
query, so a recorded run replays to identical tool results, and every
run is persisted in full: each prompt, each response, each decision and
its reason, the cards, and the final answer. What is not reproducible is
the model: under continuous batching the composition of a batch changes
the order of floating-point reductions, so two runs of one instance with
the same seed can return different text (\cref{sec:discussion:threats}).
No tool can reach the ground truth: the
gold files of the instance under investigation are not in the memory
pool, in any index, or on any path a tool will read.

\section{Evaluation Setup}
\label{sec:setup}

\subsection{Split and populations}
\label{sec:setup:protocol}
\label{sec:setup:populations}

Each project's instances are ordered by the time of their fix commit.
The earliest 30\% form the development split and the later 70\% the
evaluation split. The retrieval layer's fusion weights, its only fitted
parameters, are tuned on the development split. \sys{} has no fitted
parameters; its prompts and step budgets were settled on two small
pilots drawn from the evaluation split, which
\cref{sec:discussion:threats} discloses together with the headline
scored without those instances. We chose a time split rather than a
random one
because it mirrors deployment: a system is configured on the project's
past and then meets its future. The split has a side effect any system
with historical memory will meet: the development split holds the
oldest issues, which have the least history behind them, so the tuned
weights undervalue memory relative to its worth on later data. We keep
the split anyway; the alternative is the deployment scenario backwards.

Two populations carry the results, and no comparison mixes them. The
\emph{evaluation split} is the later 70\% of every project,
\nEval{} instances, \nEvalHard{} of them hard. \sys{}, the retrieval
layer and the loop ablation are scored on all of it; the paired
ablation of \cref{tab:ablation} uses the 26{,}537 instances both arms
answered. The
\emph{sample} is \nSample{} instances drawn from the evaluation split
with a fixed seed before any agent had run, stratified to match the
split's project mix and hard fraction (\nSampleHard{} hard against the
split's 48\%). It carries the two reimplemented published agents and
the model baselines, and every system on it, \sys{} included, is scored
on exactly those instances. The 101 benchmark instances whose gold consists entirely of
test files are excluded from both populations before any system is
scored (\cref{tab:funnel}); they are rankable and easier than average,
so the exclusion lowers every score rather than raising it.

\subsection{Metrics}
\label{sec:setup:metrics}

At the file level we report Acc@K, the
fraction of instances with at least one gold file in the top $K$, and
MRR, the mean reciprocal rank of the first gold file, with
$K \in \{1, 5\}$ and Acc@5 as the practical target, since five files is
a realistic number for an engineer to open. At the module level a
module is a file's leading path components at a fixed per-project
depth, and Acc@1 asks whether the module the system names, or the
module of its top file, contains a gold file. At the region level a
candidate is a function, identified by file, name and start line, gold
functions are those whose span overlaps a hunk of the fix diff, and
Acc@K is computed on the instances for which at least one gold function
is derivable.

\subsection{Compared systems}
\label{sec:setup:systems}

All systems see the same input, the issue report with its logs and the
code snapshot, and none sees a gold file, a memory vote or a difficulty
label. \textbf{The retrieval layer} is \cref{sec:approach}. \textbf{The
unconstrained loop} is the ablation of \cref{sec:agent:principles}: the
same model, the same evidence pass, the same tools and the same final
checks as \sys{}, with the model free to choose tools in any order for
up to fourteen turns and to decide when to submit, which is the usual
agent shape. It isolates what the fixed procedure contributes.
\textbf{LocAgent-style} and \textbf{CoSIL-style} are reimplementations
of the two published localization agents that take exactly our inputs,
an issue and a repository, and return ranked
locations~\cite{chen2025locagent,jiang2025cosil}. Neither released
system runs on this benchmark as released: both are built for Python
repositories and parse Python files only, while our projects are
mainly Java, Scala, C and C++. The graph layer
is the contribution in both papers, so replacing it is the port; we
reimplement their search strategies on our tree-sitter layer and name
the arms accordingly.
Held constant: the served model, temperature zero, the seed, the
snapshots, the reports with their logs, the context window and a turn
budget at least as large as ours. Withheld from both, because neither
design has it: incident memory, our evidence pass and our structured
submission. Both parse a free-text answer, and an instance where that
answer names no file counts as a miss. \textbf{The model alone} follows
the shape of Agentless's localization stage~\cite{xia2025agentless},
narrowing without tools: one
call picks up to fifteen directories from the snapshot's directory
list, a second ranks ten files from the source files in those
directories. \textbf{The model as re-ranker} is given the retrieval
layer's top twenty candidates with their function names and returns a
ranked ten. All model-based systems use one general-purpose
open-weights model, Qwen3.8-27B, self-hosted on H100 GPUs at
temperature zero with a fixed seed and thinking disabled, so what
separates them is the architecture.

\subsection{Statistics}
\label{sec:setup:stats}

Every comparison is paired over instances and tested with a bootstrap
of 2{,}000 resamples; we report the mean difference with its 95\%
interval and call it significant when the interval excludes zero. We do
not correct for multiple comparisons: the differences the paper builds
on are far from zero, and the one that is not, the file-level
difference between the two agents, is reported as small. The retrieval
layer is deterministic end to end and reproduces its tables bit for
bit; the agents run at temperature zero with a fixed seed, and the
residual variation of the model is reported in
\cref{sec:discussion:threats}.

\section{Results}
\label{sec:results}

Every number in this section is from the \nProjects-project benchmark.
\sys{}, the loop ablation and the retrieval layer are scored on the full
\nEval-instance evaluation split; the published agents and the model
baselines are scored on a stratified \nSample-instance sample of that
split (\cref{sec:setup:populations}). \Cref{tab:all} collects every
system and every metric in one place, one block per population; all
systems in a block are scored on identical instances. Intervals are
95\% paired bootstrap intervals over instances; ``significant'' means
the interval excludes zero.

\begin{table*}[t]
\caption{Every system on every metric. The upper block is the full
evaluation split (\nEval{} instances, \nEvalHard{} of them hard, meaning
the report names no file the fix changed); the lower block is the
stratified \nSample-instance sample on which the published agents and
the model baselines were run. All systems in a block are scored on
identical instances. Module is the module of the top-ranked file for
every row. Region metrics are on the instances whose fix diff yields a
gold function; the retrieval layer's region row is its BM25 function
stage (\cref{sec:results:regions}). ``Cites'' is the share of causal-chain
steps that cite one of the report's log events. Cost is per issue:
model calls / prompt tokens. The best value in each column of a block
is bold, and \win{green} when it is \sys{}'s.}
\label{tab:all}
\centering
\footnotesize
\setlength{\tabcolsep}{4pt}
\begin{tabular}{@{}l cccc cc c cc c c@{}}
\toprule
 & \multicolumn{4}{c}{file, all} & \multicolumn{2}{c}{file, hard} & module & \multicolumn{2}{c}{region} & chain & cost \\
\cmidrule(lr){2-5}\cmidrule(lr){6-7}\cmidrule(lr){9-10}
 & Acc@1 & Acc@5 & Acc@10 & MRR & Acc@1 & Acc@5 & Acc@1 & Acc@1 & Acc@5 & cites & calls / tokens \\
\midrule
\multicolumn{12}{@{}l}{\emph{evaluation split, 26,544 instances (12,799 hard)}} \\
retrieval layer (no model) & 41.2 & 70.3 & 78.3 & 0.539 & 20.3 & 48.7 & 76.5 & 21.9 & 42.1 & -- & 0 / 0 \\
\sys{} & \win{68.6} & \win{85.9} & \win{88.1} & \win{0.761} & \win{52.8} & \win{73.6} & \win{90.9} & \win{53.8} & \win{69.9} & \win{57.3} & 15.7 / 106k \\
\midrule
\multicolumn{12}{@{}l}{\emph{stratified sample, 1,000 instances (483 hard)}} \\
retrieval layer (no model) & 40.6 & 69.4 & 77.6 & 0.533 & 20.7 & 48.7 & 78.2 & -- & -- & -- & 0 / 0 \\
LocAgent-style & 59.9 & 71.5 & 71.5 & 0.647 & 40.4 & 52.4 & 82.8 & -- & -- & -- & 13.9 / 106k \\
CoSIL-style & 47.2 & 59.4 & 60.1 & 0.525 & 17.2 & 27.7 & 75.4 & -- & -- & -- & 18.8 / 23k \\
model alone & 52.7 & 71.0 & 76.2 & 0.607 & 29.8 & 51.6 & 87.5 & -- & -- & -- & 2.0 / 16k \\
model as re-ranker & 60.8 & 79.0 & 82.0 & 0.688 & 39.3 & 60.0 & 89.0 & -- & -- & -- & 1.0 / 3k \\
\sys{} & \win{67.1} & \win{85.1} & \win{87.5} & \win{0.751} & \win{52.2} & \win{72.0} & \win{91.5} & \win{53.3} & \win{68.7} & \win{55.7} & 15.8 / 108k \\
 
\bottomrule
\end{tabular}
\end{table*}

\subsection{What the agent reaches}
\label{sec:results:headline}

The upper block of \cref{tab:all} gives the headline. On \nEval{} real incidents
\sys{} ranks a file the fix changed first for \accOne{} of them and in
the top five for \accFive{} (MRR \mrr). The module it names is correct
for \modOne{}. At the region level, where a hit means the top-ranked
function is one the fix changed, it reaches \regOne{} Acc@1 and
\regFive{} Acc@5 on the \nRegion{} instances that have a gold
function (24{,}847 of them when paired with the loop in
\cref{tab:ablation,tab:functions}).

The hard stratum is where the difficulty of the task lies, and it is
half the benchmark. On the \nEvalHard{} reports that name no faulty
file, \sys{} reaches \hardOne{} Acc@1 and \hardFive{} Acc@5; on the easy
half it reaches 83.3\% and 97.4\%. The retrieval layer, which has no
model, reaches \retrOne{} and \retrFive{} overall and 20.3\% and
\retrHardFive{} on the hard stratum. The agent's margin over it is
$+27.4$ points of Acc@1 overall and $+32.5$ on the hard stratum: reading
the code closes most of the gap that the report leaves open.

By report type the picture is flat. Reports with a stack trace score
66.5\% Acc@1 (7{,}808 instances), reports with a log line the log layer
recognises as an event but no trace 66.6\% (5{,}739), and reports with
neither 70.8\% (12{,}993, half the split; the four unanswered instances
are not typed). The last group are
log-bearing by the source dataset's labelling but quote their evidence
in prose, an exception name or an error message inside a sentence
rather than a pasted line, and they are the easiest reports at rank
one. A trace does not make the file easier to find, because the file it
names is usually the one that detected the failure (\cref{sec:intro}).
Whether the logs themselves carry the result is not something this
split can show: no run removes the logs from the reports, and we list
that ablation as missing in \cref{sec:discussion:threats}. Across the
\nProjects{} projects Acc@1 ranges
from 54.3\% (Infinispan) to 79.6\% (Hadoop). The projects rank the same
way for the retrieval layer (Spearman $\rho = 0.87$ between the two
systems' per-project Acc@1), so project difficulty is largely a property
of the reports rather than of the agent; the per-project table is in the
artifact.

\subsection{What the structure changes}
\label{sec:results:structure}

\begin{table*}[t]
\caption{Ablation: the same model, tools and evidence pass in an
unconstrained loop against the staged workflow, on the 26{,}537
evaluation instances both arms answered. The last row is the paired
difference in points with its 95\% bootstrap interval. ``Cites'' is the share of causal-chain steps that
cite one of the report's log events.}
\label{tab:ablation}
\centering
\scriptsize
\setlength{\tabcolsep}{3pt}
\begin{tabular}{@{}l cc c ccc c c@{}}
\toprule
 & \multicolumn{2}{c}{file, all} & file, hard & \multicolumn{3}{c}{region} & chain & cost \\
\cmidrule(lr){2-3}\cmidrule(lr){5-7}
 & Acc@1 & Acc@5 & Acc@1 & Acc@1 & Acc@5 & Acc@10 & cites & calls / tokens \\
\midrule
unconstrained loop & 67.9 & 86.8 & 53.3 & 39.8 & 50.7 & 50.7 & 0.0 & 13.6 / 99k \\
\sys{} & 68.6 & 85.9 & 52.8 & 53.8 & 69.9 & 74.8 & 57.3 & 15.7 / 106k \\
\midrule
\sys{} minus loop, paired & +0.7 [+0.2, +1.2] & -0.9 [-1.2, -0.6] & -0.5 [-1.3, +0.3] & +14.1 [+13.5, +14.7] & +19.2 [+18.6, +19.9] & +24.1 [+23.5, +24.8] & +57.3 & -- \\
 
\bottomrule
\end{tabular}
\end{table*}

The unconstrained loop of \cref{sec:setup:systems} ran on the same split
with the same model and the same evidence pass; \cref{tab:ablation}
sets the two side by side. Paired on the 26{,}537
instances both completed, the staged workflow changes the file-level
result very little: Acc@1 $+0.7$ points $[+0.2, +1.2]$, Acc@5
$-0.9$ $[-1.2, -0.6]$, MRR unchanged. Both differences are significant
and both are small. The loop finds files as well as the workflow does.

What the structure changes is everything below the file. Region Acc@1
rises from \loopRegOne{} to \regOne{} ($+14.1$ points $[+13.5, +14.7]$)
and region Acc@5 from \loopRegFive{} to \regFive{} ($+19.2$
$[+18.6, +19.9]$). The two lists differ in length, and that matters for
the Acc@5 and Acc@10 columns: the loop names 3.2 regions per answer, one
function per file, while the workflow names up to three functions for
each examined file and then fills the list with keyword-ranked
functions. Region Acc@1, where list length plays no part, is the number
to read, and on the workflow's own regions alone, without the keyword
fill, it reaches 53.5\% Acc@1 and 69.4\% Acc@5. The explanation changes
in kind as well. In the loop's chains no step cites a log event; in the
workflow 57.3\% of chain steps cite one of the report's selected log
events, and 85.7\% of those citations are grounded, meaning the step
names an event id from the report or quotes a line from it word for
word, while the rest paraphrase a line and are kept, flagged. The
citation is a property of the fixed answer form, which asks for an
event id at every step, and not an emergent behaviour of the model; we
report it because it is what an engineer receives. The price is
\calls{} model calls and \tokens{} prompt
tokens per issue against the loop's 13.6 and 99k, and 286 against 199
seconds of wall time at the same concurrency.

By report type the file-level change is nil on reports with a trace
($-0.3$ $[-1.3, +0.7]$) and small but positive on reports with logs only
($+1.4$ $[+0.3, +2.4]$) and prose only ($+1.1$ $[+0.3, +1.8]$). On the
hard stratum the workflow is $0.5$ points below the loop at Acc@1
$[-1.3, +0.3]$, not significant, and $1.1$ below at Acc@5
$[-1.7, -0.5]$, while its region Acc@1 is $+13.4$ $[+12.5, +14.2]$ and
region Acc@5 $+20.9$ $[+19.9, +21.9]$ above.

\subsection{Against reimplemented published agents}
\label{sec:results:agents}

\begin{figure}[t]
\centering
\includegraphics[width=\columnwidth]{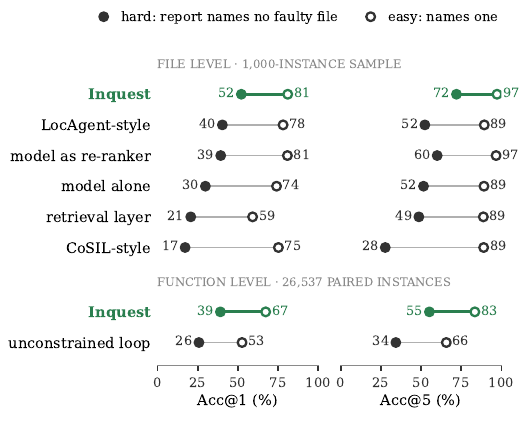}
\caption{Acc@1 and Acc@5 split by whether the report names a file
the fix changed (easy) or none (hard). Top: every system on the
\nSample-instance sample. Bottom: the two agents at the function level
on the 26{,}537 paired evaluation instances. Where the report names a
faulty file, a re-ranker matches \sys{}; where it names none, \sys{}
leads the best alternative by 12 points at Acc@1 and 72 to 60 at
Acc@5.}
\label{fig:easyhard}
\end{figure}

The lower block of \cref{tab:all} compares \sys{} with two agents from
the literature, reimplemented because neither released system runs on
Java, Scala, C or C++ (\cref{sec:setup:systems}), and with the same
model used without tools, alone and as a re-ranker of the retrieval
layer's candidates; \cref{fig:easyhard} splits every score by whether
the report names a faulty file.
\sys{} leads every system on every column. Against LocAgent-style, the
stronger published agent, the paired difference is $+7.2$ points of
Acc@1 $[+4.5, +10.0]$ and $+13.6$ of Acc@5 $[+11.0, +16.2]$; on the
hard stratum $+11.8$ $[+7.4, +16.4]$ and $+19.7$ $[+15.3, +24.0]$.
Against CoSIL-style it is $+19.9$ and $+25.7$ overall and $+35.0$ and
$+44.3$ on the hard stratum. Against the model alone it is $+14.4$
points of Acc@1 $[+11.3, +17.6]$ and $+14.1$ of Acc@5 $[+11.6, +16.7]$;
against the re-ranker, the strongest baseline at rank one, $+6.3$
$[+3.8, +8.8]$ and $+6.1$ $[+4.2, +8.2]$. Every difference is
significant. Because all five systems share the model, what separates
them is the architecture.

The cost column puts the gains in context. The re-ranker reaches 60.8\%
Acc@1 for one model call and 2.8k prompt tokens per issue, and it is a
reasonable choice where that budget is the constraint; its limit is its
candidate set, since it can only reorder the twenty files retrieval
found, and on the hard stratum it falls to 39.3\% against \sys{}'s
52.2\%. LocAgent-style spends the same tokens as \sys{} (106k against
108k) for seven points less at rank one; CoSIL-style spends a quarter
of the tokens and loses most of the hard stratum.

Two observations about the ports. LocAgent-style called its graph
traversal tool 0.09 times per run against 14.5 keyword searches, so the
multi-hop traversal that distinguishes LocAgent in its own paper goes
almost unused under this model. And both arms returned no parseable
location for 8\% to 10\% of instances (LocAgent-style 8.2\%, CoSIL-style
10.1\%), which the table scores as misses. Part of the margin is that
parsing. On the 918 instances where LocAgent-style did return a
location, it scores 65.3\% Acc@1 and 77.9\% Acc@5 against \sys{}'s
68.8\% and 86.8\% on the same instances: the paired margin is $+3.6$
points $[+0.7, +6.4]$ at rank one and $+8.9$ $[+6.5, +11.4]$ at rank
five (paired per instance before rounding), so half of the rank-one margin is parse failures and the Acc@5
margin is not. On CoSIL-style's 899 answered instances the margins are
$+15.9$ $[+12.8, +19.0]$ and $+19.9$ $[+16.9, +22.9]$ points. CoSIL's
reflective alignment step exists to repair exactly that failure, so we
expected CoSIL-style to show the lower rate; it does not, and we record
that as a property of our port we cannot explain rather than as a
finding about CoSIL. \sys{} returned an answer on every instance,
because its submission is a schema the run cannot end without.

\textbf{The model does not remember these fixes.} These are public
projects, and the model may simply have memorised their histories. We
asked it directly on the same \nSample{} instances: given only an issue
key such as KAFKA-6624 and no report text, which files did that fix
touch? It claimed to recall nothing and scored 0.0\% Acc@5. Given the
issue title as well, it claimed recall for 2.1\% of instances and scored
1.6\%. What the model-based systems contribute is inference from the
report and the code, not recitation.

\subsection{The retrieval layer, and what memory is worth}
\label{sec:results:retrieval}

\begin{table}[t]
\caption{The retrieval layer's signals and their fusion, pooled over the
\nProjects{} projects on the evaluation split (\nEval{} instances;
\nEvalHard{} hard). ``Conditioned'' is the regime-conditioned weighted
fusion of \cref{sec:approach:fusion}.}
\label{tab:retrieval}
\centering
\scriptsize
\setlength{\tabcolsep}{2.5pt}
\begin{tabular}{@{}lccc cc c@{}}
\toprule
 & \multicolumn{3}{c}{all} & \multicolumn{2}{c}{hard} & easy \\
\cmidrule(lr){2-4}\cmidrule(lr){5-6}
 & Acc@1 & Acc@5 & MRR & Acc@1 & Acc@5 & Acc@5 \\
\midrule
S1 trace mentions & 26.0 & 43.2 & 0.333 & 0.0 & 0.0 & 83.4 \\
S2 path BM25 & 19.7 & 38.6 & 0.287 & 6.0 & 17.8 & 58.0 \\
S3 content BM25 & 32.3 & 61.5 & 0.456 & 24.4 & 50.2 & 72.1 \\
S4 incident memory & 22.8 & 42.1 & 0.317 & 21.0 & 40.2 & 43.8 \\
fusion, no memory & 35.5 & 59.7 & 0.464 & 7.4 & 26.4 & 90.6 \\
fusion + memory & 38.0 & 67.1 & 0.509 & 14.4 & 41.8 & 90.6 \\
cond., no memory & 40.5 & 68.8 & 0.530 & 19.3 & 46.5 & 89.7 \\
cond.\,+\,memory (the layer) & 41.2 & 70.3 & 0.539 & 20.3 & 48.7 & 90.3 \\
 
\bottomrule
\end{tabular}
\end{table}

\sys{} uses the retrieval layer as a tool, so \cref{tab:retrieval}
reports how that layer behaves on its own across the whole benchmark.
The signals divide cleanly by stratum. Trace mentions (S1) are the best
single signal on easy reports and score exactly zero on hard ones, by
definition. Content retrieval (S3) is the best standalone signal
overall. Incident memory (S4) is the one signal whose accuracy barely
moves between the strata (43.8\% against 40.2\% Acc@5), because a past
fix can point at a file the present report never names, and on the hard
stratum it is the second-best standalone signal.

That is where memory earns its place. Adding it to unweighted fusion
lifts hard-stratum Acc@5 from 26.4\% to 41.8\%; adding it to the
conditioned fusion, which already down-weights the mention signals in
the no-mention regime, lifts it from 46.5\% to 48.7\%. The increment is
small once the weights are tuned per regime and large when they are
not, which is the practical reading: memory is cheap insurance against
a report that names nothing.

The memory signal is evaluated behind the firewall of
\cref{sec:approach:firewall}. Pooled over the projects, the
future-reports rule alone removes \firewallPct{} of the candidate votes
the index proposes, and the numbers above are what survives that cut.

\subsection{Regions and modules}
\label{sec:results:regions}

\begin{table}[t]
\caption{Function-level results on the evaluation instances whose fix
diff yields at least one gold function: the BM25 stage on the instances
the retrieval layer ranked, the two agents on the instances both
answered. ``Oracle'' is the share whose
gold function is anywhere in the candidate set; it is the ceiling of
the BM25 stage and is not defined for the agents, which choose their
own regions.}
\label{tab:functions}
\centering
\scriptsize
\setlength{\tabcolsep}{2.5pt}
\begin{tabular}{@{}lcccccc@{}}
\toprule
 & eligible & Acc@1 & Acc@5 & Acc@10 & MRR & oracle \\
\midrule
BM25 over the layer's top-10 files & 24,820 & 21.9 & 42.1 & 51.1 & 0.313 & 77.0 \\
unconstrained loop (ablation) & 24,847 & 39.8 & 50.7 & 50.7 & 0.446 & -- \\
\sys{} & 24,847 & 53.8 & 69.9 & 74.8 & 0.600 & -- \\
 
\bottomrule
\end{tabular}
\end{table}

Engineers need code regions, not files. Gold functions are
derived mechanically: tree-sitter lists every function span in the
snapshot, and a function is gold when its span overlaps a hunk of the
fix diff; a gold function exists for 93.6\% of the evaluation
instances (24,850 of the 26,540 \sys{} answered), and the function
stage ranks 26,442 instances, 24,820 of them with a gold function.
\Cref{tab:functions}
compares three ways of getting there. A BM25 stage that ranks the
function bodies in the retrieval layer's top ten files reaches 21.9\%
Acc@1 with a 77.0\% ceiling, so its problem is ranking inside large
files, not finding them. The unconstrained loop reaches 39.8\%, and \sys{}
53.8\%, because its \textsc{examine} step reads each candidate file
whole, or in whole functions, and names the functions that hold the
fault while the code is on screen. On the hard stratum \sys{} reaches
39.2\% region Acc@1 and 55.2\% Acc@5.

At the module level, where modules are the leading path components of a
file at a fixed per-project depth, \sys{} names the correct module for
\modOne{} of instances (85.5\% hard, 94.6\% easy), and the module of its
top-ranked file is correct for 90.9\%. The retrieval layer's top file
gives \moduleRetr{}. Module spaces are small and defects concentrate, so
these numbers need the sanity baselines of \cref{tab:modules}: ranking
the modules that exist in the snapshot by their defect frequency on the
development split already reaches \moduleFreq{} Acc@1, and the agent's
margin over that prior, 41 points, is what counts as localization.

\begin{table}[t]
\caption{Module level on the evaluation split, with two sanity
baselines: the expected accuracy of a random module ranking
(analytical, per snapshot) and ranking the snapshot's modules by their
defect frequency on the development split.}
\label{tab:modules}
\centering
\footnotesize
\begin{tabular}{@{}lcc@{}}
\toprule
 & Acc@1 & Acc@5 \\
\midrule
random module ranking (expected) & 5.7 & 25.8 \\
development-split defect frequency & 49.3 & 76.8 \\
\midrule
\retr{} (module of the top file) & 76.5 & 96.7 \\
unconstrained loop (module of the top file) & 90.8 & -- \\
\sys{} (module named by the agent) & 90.2 & -- \\
 
\bottomrule
\end{tabular}
\end{table}

\subsection{Cost, and a dense alternative}
\label{sec:results:cost}

\sys{} spends \calls{} model calls and \tokens{} prompt tokens per
issue and 286 seconds of wall time at 32 concurrent issues on one H100,
which came to about six issues per minute per GPU over the run. The
retrieval layer answers from one CPU core with no model, in a median of
under 0.2 seconds per issue on every project. A team that can host a
27B model gains $+27.4$ points of Acc@1 for that compute; a team that
cannot, because source and production logs may not leave the building,
still has a pipeline with no GPU, no training and no network that puts
a changed file in the top five for seven issues in ten.

Log-to-code matching is often framed as a semantic retrieval problem,
so we also measured what a small dense retriever buys inside the
retrieval layer. On
200-instance samples per project, replacing BM25 with bge-small
embeddings in the content slot loses on every one of the \nProjects{}
projects (58.4\% to 44.9\% mean Acc@5): matching a log line to code is
mostly identifier matching, and the embedder reads at most 1,400 words
of a file where BM25 sees all of it. In the incident-memory slot, where
the task is report-to-report similarity, dense retrieval wins on 22 of
27 projects (37.6\% to 40.4\%) and lifts the full system from 67.7\% to
69.1\% with weights tuned for the lexical signal. Semantics pays in the
memory slot first; the code slot needs the model to read the code, which
is what the agent does.

\subsection{What goes against us}
\label{sec:results:negatives}

Five things. First, the staged workflow is below the unconstrained loop at
Acc@5 by 0.9 points, significantly, and gains only 0.7 at Acc@1: at the
file level the structure buys nothing, and a reader who wants files only
should take the loop. Second, on reports with a stack trace the
workflow is not better at all. Third, four of \nEval{} instances
ended without an answer in two runs each (the final step's output
failed to parse) and are scored as misses. Fourth, the port check for CoSIL-style fails,
as stated above. Fifth, the module the agent names (\modOne{}) is below
the module of its own top-ranked file (90.9\%) and of the loop's
(90.8\%): naming the module is a separate decision in the rank step,
and it is the weaker of the two.

\section{Discussion and Threats to Validity}
\label{sec:discussion}

\subsection{Three difficulties, measured}
\label{sec:discussion:difficulties}

\textbf{Root cause beyond symptoms.} This is the hardest part of the
task, and the benchmark shows it in a form that can be counted:
in \nHardPct{} of real reports no file the fix changed is named
anywhere in the text, and a method that matches the report against the
code it mentions scores zero there by construction. On those
\nEvalHard{} evaluation instances the retrieval layer reaches
\retrHardFive{} Acc@5 and \sys{} \hardFive{}. Two things recover a file
no log line names: incident memory, statistically, since a past fix
points at code the report never mentions; and reading the code,
causally, since the agent judges whether a file creates the bad state
or only reports it and walks back to the callers. The walk-back is one hop, not data-flow analysis. At \hardFive{}
the hard stratum is where future work should be judged, and it is not
solved.

\textbf{Noisy and incomplete logs.} Here the benchmark is kinder than
production: its logs are the excerpts reporters chose to paste, so a
human has filtered the noise once, and 30\% of reports carry a stack
trace. The pipeline groups repeated lines into events, ranks and labels
them, and finds the code that printed each by matching it against the
snapshot's logging statements; the agent cites those event ids. We did
not evaluate on synthetic noise or raw runtime bundles, which would
need segmenting by incident first.

\textbf{Evidence-to-code semantic retrieval.} Our results say where
log-and-code semantics have to sit. A small dense
retriever helps in the memory slot, where the task is comparing one
report with another, and loses to full-text BM25 in the code slot on
every one of the \nProjects{} projects, because matching a log line to
code is mostly identifier matching (\cref{sec:results:cost}). What
supplies the semantics in our system is a model that reads the code,
paired with a procedure that puts the right code in front of it.
The lower block of \cref{tab:all} is a ladder of how much code the same
model sees: path names only, 52.7\% Acc@1; paths with function names,
60.8\%; the files themselves, read whole or in whole functions, 67.1\%.

\subsection{What the explanation output is}
\label{sec:discussion:explanation}

The retrieval layer produces provenance rather than prose: for each
ranked file, which signals ranked it, the matching terms, and the past
incidents that voted for it. \sys{} produces an argument: for each file, the verdict and
the quoted line it rests on, and one chain from the reported symptom to
the named cause in which 57.3\% of steps cite one of the report's log
events, 85.7\% of them grounded, as an event id or a verbatim quote.
Both are checkable without being
trusted: a chain can cite a real line and still give the wrong cause,
so the evidence is a lead to verify, and no user study shows that
either form speeds engineers up.

\subsection{Threats to validity}
\label{sec:discussion:threats}

\textbf{Construct.} The answer key is the set of non-test files a fix
modified, a proxy for causation rather than a measurement of it. A
commit subject can name an issue it does not fix (reverts, merges and
long multi-commit efforts are removed); a fix can touch files
refactored in passing; and the repaired file is occasionally not where
the defect originated. With a
median of one gold file per instance the effect is bounded; accuracy
rises with fix size for every system because a larger fix offers more
files to hit, a loosening of the key rather than better localization. The module metric depends on a per-project depth
fixed before evaluation. Gold functions are derived from diff hunks and
miss a defect fixed by adding a function.

\textbf{Internal.} The prompts and step budgets of \sys{} (the
candidate table of thirty, the five files to shortlist, the eight to
examine, the 24{,}000-character file view) were developed on two
pilots drawn from the evaluation split, of 30 and 81 instances, both
disjoint from the \nSample-instance sample; no parameter was fitted to outcomes on evaluation instances, and
the pilot instances remain in the split (0.4\% of it). Scoring \sys{}
on the 26{,}429 answered instances outside the pilots changes nothing at the
paper's precision: file Acc@1 68.6\% and Acc@5 85.9\%, module 90.2\%,
region 53.8\% and 70.0\%. Two ablations
are missing. No run removes the logs from the reports, so the
contribution of the log lines themselves, as opposed to the rest of
the report, is not isolated; and incident memory is ablated in the
retrieval layer (\cref{tab:retrieval}) but not inside the agent, so
what the agent's \texttt{recall} evidence contributes on its own is not
measured. The agents run at temperature zero with a fixed seed, but
the model is not bit-reproducible under continuous batching.
On 20 instances run twice, the top-ranked file agreed on 18 on the same
machine and on 14 to 15 across two machines with identical software and
model checksums, and the module on all 20: the top file agrees across
repeats in 70\% to 90\% of cases, so differences of a point or two
between agent runs are inside the noise. Three seeds of \sys{} on
the \nSample-instance sample bound that noise in the aggregate: on the
998 instances all three answered, file Acc@1 spans 67.0\% to 67.9\%,
module 89.4\% to 90.2\% and region Acc@1 53.2\% to 53.3\%, at most
one point on every headline metric, while 55 to 61 instances change
their top file between two seeds. Four instances ended without an answer and are
scored as misses. Thinking was disabled for every model-based system, a
fair comparison that may understate the model.

\textbf{External.} Twenty-seven projects with disciplined issue-commit
linkage, mainly in four languages; a project without such linkage
cannot use the mining recipe directly. One model, so the result shows
what this architecture achieves with a 27B open-weights model, not how
it scales. Logs are
reporter-curated, as discussed above.

\textbf{Baselines.} LocAgent-style and CoSIL-style are our
reimplementations, not the published systems, and one check on the
ports goes the wrong way: CoSIL-style, whose design includes a step to
repair unparseable answers, left more answers unparseable than
LocAgent-style. Both ports leave 8\% to 10\% of instances without a
parseable answer, which counts as a miss; \cref{sec:results:agents}
reports the margins with those instances excluded as well. Pathidea,
the closest log-based prior work, was not run, not even on the Java
subset its path reconstruction is built for; that is a gap. Classic information-retrieval
localizers were not run; coverage-based localization needs execution
evidence absent at report time. The contamination probe
tests recitation from an issue key, not recognition when report and
code are shown, and the benchmark has no post-cutoff subset, so
memorisation cannot be excluded. The explanations were checked mechanically, for existence
and grounding; no human judged whether a chain's reasoning is correct,
and no practitioner has used the output.

\section{Related Work}
\label{sec:related}

\textbf{Localization that needs execution.} Spectrum-based and
mutation-based fault localization rank code by how it behaves under
failing and passing tests~\cite{wong2016survey}, and recent
agent-based variants keep that dependency, using coverage or a
reproduction test~\cite{rafi2024multiagent,yu2025orcaloca}. The
evidence they need does not exist when an issue is filed: on crash
reports, only 2 of 60 bugs had a fault-triggering test available at
report time~\cite{rafi2025sbest}. Our setting excludes execution
evidence by construction.

\textbf{Information-retrieval bug localization.} The BugLocator lineage
ranks files by textual similarity to the report and adds SimiScore, a
vote from similar past reports~\cite{zhou2012buglocator}; BRTracer adds
stack-trace boosting~\cite{wong2014brtracer}; later work adds learned
rankers~\cite{ye2014learning} and deep variants~\cite{niu2025deepirbl}.
Bench4BL established the reproduction baseline~\cite{lee2018bench4bl},
and BLAZE extends it across languages~\cite{chakraborty2025blaze}.
SimiScore is the same mechanism as our incident memory, and our
firewall is what that mechanism needs once report-time admissibility
is taken seriously. Pathidea is the closest prior work on logs: it
scores files using the traces and log lines in the report and
reconstructs execution paths between log statements, on Java
only~\cite{chen2022pathidea}; its log score informs our trace signal.

\textbf{LLM and agent localization.} Agentless showed that a two-stage
procedure over the repository tree is a strong
baseline~\cite{xia2025agentless}, and it is our model-alone arm. Agents
navigate structure: LocAgent over a code graph with an entity
index~\cite{chen2025locagent}, CoSIL over dynamically built call
graphs~\cite{jiang2025cosil}, GenLoc by seeding an LLM with retrieval
candidates and letting it explore~\cite{asad2026genloc}. \sys{} shares
GenLoc's seed-then-explore shape and adds the project's own resolved
history behind a temporal firewall and a procedure fixed in code,
measured against the unconstrained loop these systems use. CoSIL's
analysis of LLM localizer failures, malformed tool arguments and
hallucinated paths, matches what we saw and shaped our fixed answer
forms.

\textbf{Incident history in operations.} Retrieving similar past
incidents is established practice in cloud operations. RCACopilot
retrieves prior incidents by embedding similarity with time decay and
reports that most recurring incidents reappear within
days~\cite{chen2024rcacopilot}; RCAgent performs tool-augmented
root-cause analysis~\cite{wang2024rcagent}; LLM-based recommendation
of root causes is deployed at scale~\cite{ahmed2023rootcause}. These
systems identify a faulty service or produce prose and stop above the
level of source files, which is the step this paper takes.

\textbf{Benchmarks, leakage and contamination.} Benchmarks built from
public repositories leak solutions~\cite{aleithan2024swebenchplus},
models reproduce file paths better for repositories they were trained
on~\cite{liang2025illusion}, a replicated localizer's reported advantage
came from history that post-dates the report~\cite{niu2026ablots}, the
best agents lose most of their accuracy on a newer and broader issue
set~\cite{zhang2025mulocbench}, and time-aware evaluation changes
which method appears best~\cite{ledel2021broccoli}. We therefore use a
time-based split and a four-rule firewall. An industrial evaluation of
bug localization found simple term-frequency methods competitive and
reported the constraints that dominate adoption~\cite{hall2026industrial};
our cost reporting and our CPU-only fallback follow the same concern.

\section{Conclusion}
\label{sec:conclusion}

This paper presented a benchmark and an agent for log-guided root-cause
localization. The benchmark pairs \nInstances{} real incidents from
\nProjects{} projects with their logs, the code while the defect was
present, and the files the fix changed; half of its reports name
no faulty file. \sys{} localizes those incidents by a fixed
procedure in which the model decides only which files to open, what
role each plays, and their order. On \nEval{} evaluation instances it
puts a changed file first for \accOne{} and in the top five for
\accFive{}, names the module for \modOne{} and the function for
\regOne{}.

Two results stand on their own.
An unconstrained agent loop with the same model and tools finds files
as well as the staged procedure does: the two tie at the file level on
26{,}537 paired instances.
What structure buys is below the file, fourteen points of
function-level accuracy and an explanation whose steps cite the
report's own log events. And incident memory, admitted only when it would have existed at report
time, is the one retrieval signal whose accuracy does not collapse when
the report names nothing. The benchmark\footnote{\url{https://doi.org/10.5281/zenodo.24217071}}, both
systems, every run transcript and the script that ties each number
here to its result file\footnote{\url{https://doi.org/10.5281/zenodo.24217136}}
are released.

\bibliographystyle{IEEEtran}
\bibliography{refs}

\end{document}